\documentclass[a4paper,twocolumn]{article}
\usepackage[utf8]{inputenc}
\usepackage{hyperref}
\usepackage{enumitem}
\usepackage{csquotes}
\MakeOuterQuote{"}

\title{Leaky Abstraction In Online Experimentation Platforms: A Conceptual Framework To Categorize Common Challenges}
\author{Timo Kluck \\ timo.kluck@booking.com 
	\and Lukas Vermeer \\ lukas.vermeer@booking.com}
\date{}

\begin{document}

\maketitle

\begin{abstract}

Online experimentation platforms abstract away many of the details of experimental design, ensuring experimenters do not have to worry about sampling, randomisation, subject tracking, data collection, metric definition and interpretation of results. The recent success and rapid adoption of these platforms in the industry might in part be attributed to the ease-of-use these abstractions provide \cite{Bakshy14}. Previous authors have pointed out there are common pitfalls \cite{Crook09, Kohavi09, Kohavi12} to avoid when running controlled experiments on the web and emphasised the need for experts familiar with the entire software stack to be involved in the process \cite{Kohavi12, Tang10}.

In this paper, we argue that these pitfalls and the need to understand the underlying complexity are not the result of shortcomings specific to existing platforms which might be solved by better platform design. We postulate that they are a direct consequence of what is commonly referred to as "the law of leaky abstractions" \cite{Spolsky02}. That is, it is an inherent feature of any software platform that details of its implementation leak to the surface, and that in certain situations, the platform's consumers necessarily need to understand details of underlying systems in order to make proficient use of it.

We present several examples of this concept, including examples from literature, and suggest some possible mitigation strategies that can be employed to reduce the impact of abstraction leakage. The conceptual framework put forward in this paper allows us to explicitly categorize experimentation pitfalls in terms of which specific abstraction is leaking, thereby aiding implementers and users of these platforms to better understand and tackle the challenges they face.

\end{abstract}

\section*{Introduction}

\subsection*{The law of leaky abstractions}
Software platforms invariably expose an abstract API to hide certain complexity from their users. However, it can be argued that they can never completely succeed in concealing the intricacies of the underlying subsystems and that computer abstractions are therefore always imperfect. This phenomenon was first described by Kiczales \cite{Kiczales92}, and the terminology "law of leaky abstractions" was introduced by Spolsky \cite{Spolsky02}. As an example, virtually every function call in Python might raise a \texttt{MemoryError} exception, yet this is usually not part of their specification. The common abstraction is one of infinite memory. Another example is that different, yet logically equivalent, SQL queries may have very different performance characteristics \cite{Spolsky02}.

As a corollary, users of such computer abstractions can, and do, run into situations where understanding the abstraction does not suffice for attaining the desired results. A Python developer who is confronted with \texttt{MemoryError} exceptions needs to understand \emph{how} their function call is allocating memory, and possibly how to decrease these allocations. An SQL developer who is confronted with slow-running queries needs to understand the inner workings of their database's query optimizer to understand how to obtain the desired performance.

\subsection*{Abstractions in online experimentation platforms}
Abstraction in software systems is not limited to hiding complexities of underlying software or hardware systems. In many cases, it applies to theoretical concepts or even objects in physical space. Experimentation platforms in particular commonly abstract away the following aspects of experimental design:

\begin{description}[style=unboxed,leftmargin=0cm]

\item [Statistical unit and tracking.] The platform determines what is the statistical unit of observation (e.g. login account, cookie or uninterrupted sessions). Then for a given event (e.g. http request) it finds the corresponding unit and tracks accordingly.

\item [Sampling.] The platform imposes treatment to a (usually randomized) subset of traffic. It takes responsibility for ensuring that this selection of current traffic is representative of traffic as a whole. When reporting the resulting metrics, and showing them with confidence intervals and predictions of overall impact, it is implicitly taking responsibility for the assumption that both the treatment and control selections of current traffic are representative of future traffic. It also assumes that on the side of the IT infrastructure, showing treatment to a percentage of traffic has the same (e.g. performance) characteristics as showing treatment to full traffic.

\item [Definition and implementation of metrics.] The platform defines how metrics are gathered and thereby also their exact definition.

\item [Business meaning of metrics.] The platform will likely interpret an improvement in click-through rate, revenue, or conversion as being "good". It may highlight this result in a positive way, or even end the experiment and select the treatment as the new default.

\end{description}

We observe that all four of these abstractions can leak, in which case the experimenter needs to be well-informed of their intricacies in order to base a sensible decision on the collected data. This has been alluded to by Tang et al. \cite{Tang10}

\begin{displayquote}

There are times (\ldots) where something in the actual implementation goes awry, or something unexpected happens. In those cases, the discussion is as much a debugging session as anything else. Having experts familiar with the entire stack of binaries, logging, experiment infrastructure, metrics, and analytical tools is key.

\end{displayquote}

The conceptual framework put forward in this paper can help explain why the need for "experts familiar with the entire stack" can not be avoided, as well as enable us to categorize common pitfalls in terms of which specific abstraction is leaking.

\section*{Examples of leaky abstraction}

Let us consider a few known experimentation pitfalls through the lens of leaky abstraction. Kohavi et al. \cite{Kohavi12} describe a situation where a certain experiment showed an unexpected but statistically convincing uplift in click-through rate.

\begin{displayquote}

The "success" of getting users to click more was not real, but rather an instrumentation difference. Chrome, Firefox, and Safari are aggressive about terminating requests on navigation away from the current page and a non-negligible percentage of click-beacons never make it to the server. (\dots) Adding even a small delay [in the treatment] gives the beacon more time, and hence more click request beacons reach the server.

\end{displayquote}

We can regard this situation as the experiment framework implementing the metric of "click-through rate" as "successful request beacons reaching the server". This implementation is abstracted away, but in this particular instance, it leaks to the surface. Knowing that this is a situation where this abstraction is known to fail, we could mitigate the future impact of leaks by warning experiment implementers when using beacons close to a point where the user is likely to initiate navigation.

A key point in this previous example is that \emph{it is the treatment itself that interferes with the platform's workings}. This means that implementing the feature and measuring its impact in an experiment are not concerns that can be separated. The experimenter needs to understand the platform's implementation.

As another example, consider the situation where treatment involves caching certain data for each user. One might imagine a situation where 50\% of traffic will have caching needs that can be satisfied by the infrastructure with a very low eviction rate, whereas 100\% of traffic would lead to a much higher eviction rate and consequently worse performance. This would likely result in abstraction leakage, as the experimentation platform silently assumes that exposing 50\% of traffic to treatment will have the same effect as exposing 100\% of traffic to it.

In this case, there is \emph{nothing the experimentation platform could do} to flag this issue. The experimenter has to rely on their own experience and analysis to foresee this possible problem. A possible mitigation strategy would be to also test the experiment at different traffic splits (e.g. if the treatment is positive for 50\% of traffic, then run the experiment again for 90\% of traffic). In this particular example, one can also carefully monitor cache eviction rates.

\section*{Mitigating abstraction leakage}

The law of leaky abstractions states that leakage is unavoidable. Nonetheless, we propose three general methods for reducing the risks:

\begin{description}[style=unboxed,leftmargin=0cm]

\item [Increased user awareness.] Users of experimentation platforms must be made conscious of the abstractions the platform is hiding from them. Personal experience tells us that users are often not aware of the design decisions that were made by platform developers. By making these chosen defaults explicit, users are in a better position to design experiments that work with the platform rather than against it.

\item [Expert experiment review.] Expert review and assistance setting up experiments should be made available when needed. A formal process or some additional user education may be required to help users identify potentially problematic scenarios.

\item [Automated warnings for known pitfalls.] Automated warnings for known pitfalls should be implemented as part of the platform user interface. This can flag potential issues for more in-depth review; possibly aided by experts. Alerting users to potential issues and guiding them in their analysis of the underlying complexity has the added benefit of serving as additional “training on the job”, which might help experimenters better identify future issues which the system does not detect.

\end{description}

\section*{Conclusion}

In this paper we have identified four aspects of experimental design commonly abstracted away by experimentation platforms: statistical unit and tracking, sampling, definition and implementation of metrics and business meaning of metrics. Two examples of abstraction leakage are presented: click beacons and cache eviction. Finally, we suggest how to mitigate these through increased user awareness, expert experiment review and automated warnings for known pitfalls.

The conceptual framework provided in this paper aids implementers and users of online experimentation platforms to better understand and tackle the challenges they face. Leaky abstraction can function as a helpful lens through which to consider several previously known pitfalls. 

\section*{Acknowledgements}

The ideas put forward in this paper were greatly influenced by our work on the in-house experimentation platform at Booking.com, as well as conversations with colleagues and other online experimentation practitioners.

\end{document}